\documentclass[12pt]{article}
\usepackage{dcolumn}
\usepackage{bm}
\usepackage{amsmath}
\usepackage{amssymb}
\usepackage{amscd}
\usepackage{afterpage}
\usepackage{float,times}
\usepackage{subfigure}
\usepackage{rotating}
\usepackage{multirow}
\usepackage{epsfig}
\usepackage{theorem}
\usepackage{moreverb}
\usepackage{euscript}
\usepackage{psfrag}

\begin{document}

\begin{titlepage}
\begin{center}
{\hbox to\hsize{\hfill May 2009 }}

\bigskip

\bigskip

\bigskip
\vspace{6\baselineskip}

{\Large \bf
Diffeomorphism-invariant noncommutative gravity   \\ with twisted local Lorentz invariance \\
}
\bigskip

\bigskip

{\bf 
 Archil Kobakhidze  \\}
\smallskip

{ \small \it 
School of Physics, The University of Melbourne, Victoria 3010, Australia \\ 
E-mail: archilk@unimelb.edu.au \\}

\bigskip

\vspace*{.5cm}

{\bf Abstract}\\
\end{center}

\baselineskip=13pt

\noindent
{\small 
We propose a new theory of gravitation on noncommutative space-time which is invariant under the general coordinate transformations, while the local Lorentz invariance is realized  as twisted gauge symmetry. Our theory is remarkably simpler compared to the existing formulations of noncommutative gravity.  }  

\bigskip

\bigskip

\bigskip

\end{titlepage}

\baselineskip=13pt

\section{Introduction}

It is conceivable to think that classical singularities in General Relativity and quantum ultraviolet divergences in quantum field theory both are the artifacts of the accepted concept of smooth differentiable manifolds for space-time. The idea that one must instead consider some sort of quantized space-times traces back to Heisenberg and Pauli, with the first published paper on the subject written by Snyder in 1947 \cite{Snyder:1946qz}.  The most elegant way to view such quantized space-times is to assume that space-time coordinates do not commute.  Despite many important developments it is fair to say that physical implications of space-time noncommutativity are still less explored. 

In this regard formulation of noncommutative theory of gravitation is one of the most interesting and challenging tasks. In principle, such theory of gravitation could be constructed based on the powerful mathematical theory of noncommutative differential geometry developed by Connes \cite{Connes:1994yd}. In practice, however, only limited types of noncommutative spaces are known how to treat explicitly in this framework \cite{Chamseddine:1995xh}. More recently, several attempts have been made to formulate theory of gravitation on so-called canonical noncommutative space-time. The key problems one usually faces are a proper implementation of symmetries of General Relativity and an explicit construction of the connections on quantum manifolds. These formulations essentially differ in the way these problems are solved. In \cite{Aschieri:2005yw} , \cite{Kobakhidze:2006kb} the symmetries of General Relativity are replaced by their twisted counterparts. The physical relevance of twisted diffeomorphism invariance of \cite{Aschieri:2005yw} has been doubted in \cite{Alvarez-Gaume:2006bn}. Also the twisted approach to gauge symmetries \cite{Vassilevich:2006tc} used in \cite{Kobakhidze:2006kb} has been criticized in \cite{Chaichian:2006we}, \cite{Chaichian:2006wt}. The key drawback is that the $\star$-product defined in these works is not covariant. An attempt to construct theory of gravitation with by gauging twisted Poincare group \cite{Chaichian:2004za} with the covariant $\star$-product \cite{Chaichian:2006wt} has been made in \cite{Chaichian:2008pq}. However, the resulting theory turns out to be nonassociative. Alternatively, in \cite{Calmet:2005qm} noncommutative gravity is constructed based on symplectic volume-preserving diffeomorphism, which is the "true" symmetry of canonical noncommutative spacetimes (see also \cite{Banerjee:2007th} for analogous construction in Lie-algebraic noncommutative space-time). Some attempts to construct noncommutative gravity with space-time dependent noncommutativity parameter has been made in \cite{Harikumar:2006xf}, \cite{Marculescu:2008gw}, while in \cite{Yang:2004vd} theories of emergent noncommutative gravity has been suggested.

\section{Symmetries of General Relativity} 

In this section we would like to recollect well known facts about General Relativity. The General Relativity can be viewed as a gauge theory described by two separate symmetry groups of local transformations: one is the group of General Coordinate Invariance (GCI), and another is the group of Local Lorentz Invariance (LLI). An arbitrary element of the combined group can be represented as, 
\begin{equation}
g(x)=e^{i\xi^{\mu}P_{\mu}+\frac{i}{2}\lambda^{ab}(x)\Sigma_{ab}}~.
\label{1}
\end{equation}  
The first term in the exponential describes general transformations\footnote{More precisely, these are the transformations which approach unity at infinity.} of space-tme coordinates $x^{\mu}$ generated by the  generators $P_{\mu}=-i\partial_{\mu}$. The second term in (\ref{1}) describes Lorentz rotations of inertial coordinates $\xi^a_{x}(y)$ defined locally in the infinitesimal vicinity of a given point $y^{\mu}=x^{\mu}$ of the space-time manifold, with  $\Sigma_{ab}$ being the standard Lorentz generators.  Note that the $P_{\mu}$ and $\Sigma_{ab}$ act in different spaces, so that they commute, $[P_{\mu}, \Sigma_{ab}]=0$.   

To gauge the above group we introduce the gauge potential, 
\begin{equation}
A_a= e^{\mu}_aP_{\mu}+\omega_{a}^{~cd}\Sigma_{dc}~, 
\label{2}
\end{equation} 
where the (inverse) vierbein $e^{\mu}_{a}(x)$ is a gauge field of GCI group, while the spin connection  $\omega_{\mu}^{~cd}(x)=e^{a}_{\mu}\omega_{a}^{~cd}$ is the gauge field for LLI group. On the other hand, from the geometric point of view, General Relativity can be considered as a theory of (pseudo)Riemannian manifolds. As it is well-known, in the vicinity of any given point $x$ of the Riemannian manifold we can define locally inertial coordinates $\xi_{(x)}^a$, in which  $e_{\mu}^{a}=\delta^{a}_{\mu}$ and $\omega_{\mu}^{~ab}=0$. This is nothing but the statement of the {\it Equivalence Principle}: locally there exist coordinates in which the effects of gravitation are absent. 

\section{Covariant noncommutative space-time}

Let us now define a covariant noncommutative space-time. It seems that the most natural generalization of the canonical flat space-time noncommutativity to the curved space-time is to assume canonical noncommutativity of a tangent space-time, i.e.
\begin{equation}
[\hat \xi^a, \hat \xi^b]=i\ell^2\theta^{ab}~,
\label{5}
\end{equation}
where $\hat \xi^a$ are noncommutative tangent space coordinates, $\ell$ is the length scale defining noncommutativity, and $\theta^{ab}=-\theta^{ba}$ are constants. We stress that $\theta^{ab}$ represent just a collection of constants rather than SO(3.1) tensor, that is to say, it does not transform under the SO(3.1) transformations. This implies, in particular, 
\begin{equation}
(A_{a}\theta^{bc})=e_a^{~\mu}\partial_{\mu}\theta^{bc}=0~.
\label{5a}
\end{equation}

The algebra of noncommutative coordinates given by (\ref{5}) can be mapped onto the $\star$-algebra of commutative coordinates, $\xi^a$, through the suitably choosen $\star$-multiplication. We demand that this $\star$-product is covariant both under the GCI and (twisted, see below ) LLI transformations. That is to say, it involves covariant derivatives, 
\begin{equation}
\star \stackrel{\rm def}{=}{\rm exp}\left(-\frac{i\ell^2}{2}\stackrel{\leftarrow}{A_a} \theta^{ab}\stackrel{\rightarrow}{A_b}\right)~.
\label{6}
\end{equation}
Remarkably, given the above $\star$-product (\ref{6}) and taking into account the metricity condition, $(A_{a}e_{b}^{~\mu})=0$, we find that the the vierbein field commute with an arbitrary fields $f(x)$, 
\begin{equation}
[e^{~\mu}_a(x), f(x)]_{\star}=e^{~\mu}_a(x)\star f(x)-f(x)\star e^{~\mu}_a(x)=0~,
\label{7}
\end{equation} 
as in the usual commutative gravity. Since in the locally inertial coordinate system the covariant derivatives $A_a$ act as an ordinary derivatives, it is obvious that the albebra (\ref{5}) indeed can be realized as a $\star$-commutator of commutative coordinates, 
\begin{equation}
[\xi^a,\xi^b]_{\star}\stackrel{\rm def}{=}\xi^a\star \xi^b - \xi^b \star \xi^a=i\theta^{ab}~,
\label{8}
\end{equation}
while generic space-time coordinates satisfy the following commutation relations,
\begin{equation}
[x^{\mu}, x^{\nu}]_{\star}=i\theta^{\mu\nu}(x)\stackrel{\rm def}{=}ie^{~\mu}_a(x)e^{~\nu}_b(x)\theta^{ab}
\label{9}
\end{equation}
Notice that $\theta^{\mu\nu}(x)$, as it is defined in the above equation, transforms nontrivially under the local Lorentz transformations, and it also transforms under the GCI group of transformations as a 2nd rank tensor\footnote{Similar space-time dependent noncommutative tensor has been considered in \cite{Aschieri:2008zv} within the noncommutative scalar field theory.}. It is not hard to realize that the noncommutative parameters $\theta^{\mu\nu}(x)$ also commute with an arbitrary space-time function, 
\begin{equation}
[f(x), \theta^{\mu\nu}(x)]_{\star}=0~.
\label{10}
\end{equation}
From (\ref{10}) it follows that the algebra defined by (\ref{9}) is associative,
\begin{equation}
[x^{\mu},[x^{\nu},x^{\rho}]_{\star}]_{\star}+\mathrm{cyclic~ permutations}= i[x^{\mu}, \theta^{\nu\rho}(x)]_{\star}+\mathrm{cyclic~ permutations}=0
\label{10a}
\end{equation}

The above construction of noncommutative space-time is natural from the point of view of \emph{Noncommutative Equivalence Principle}: locally there exists a coordinate system where the effects of noncommutative gravitation are absent and the space-time looks as the flat noncommutative space-time with canonical noncommutativity. 

\section{Symmetries of noncommutative General Relativity}

Obviously LLI can not be realized on noncommutative fields with $\star$-algebra defined by the product (\ref{6}), since it is explicitly broken in (\ref{5}) and also in (\ref{9}) However, using the suitable twist, LLI can be shown to be a symmetry compatible with $\star$-multiplication. Consider, for example, two noncommutative fields $\Phi_1(x)$ and $\Phi_2(x)$ transforming under the GGI and LLI as,
\begin{equation}
\delta\Phi_1 = g_1\Phi_1~,~~\delta\Phi_2 = g_2\Phi_2~.
\label{11}
\end{equation} 
In order to define how the product of fields are transformed under the twisted LLI symmetry we first introduce a twist operator, 
\begin{equation}
{\cal T}={\rm exp}\left(-\frac{i\ell^2}{2}\stackrel{\leftarrow}{A_a} \theta^{ab}\stackrel{\rightarrow}{A_b}\right)~.
\label{12}
\end{equation}
Using this twist operator the $\star$-product map $\mu_{\star}$ can be writtent in terms of the usual pointwise product map $\mu_{0}$\footnote{ $\mu_{0}\left(\Phi_1(x)\otimes \Phi_2(x)\right)\stackrel{\rm def}{=}\Phi_1(x)\Phi_2(x)$, $\otimes$ denotes the tensor product.} as, 
\begin{equation}
\Phi_{1}(x)\star \Phi_2(x)\stackrel{\rm def}{=}\mu_{\star}\left\{\Phi_1(x)\otimes \Phi_2(x)\right\}=\mu_{0}\left\{{\cal T}\Phi_1(x)\otimes \Phi_2(x)\right\}~.
\label{13}
\end{equation} 
We say that LLI is realized as a symmetry on the $\star$-algebra of functions if the following compatibility condition holds, 
\begin{equation}
\mu_{\star}\left\{\delta\left(\Phi_1(x)\otimes \Phi_2(x)\right)\right\}~,
\label{14}
\end{equation}
where the transformation of the product of fields is defined through the so-called co-product $\Delta_{\star}(\delta)$, 
\begin{equation}
\delta\left(\Phi_1(x)\otimes \Phi_2(x)\right)\stackrel{\rm def}{=}\Delta_{\star}(\delta)\left(\Phi_1(x)\otimes \Phi_2(x)\right)~.
\label{15}
\end{equation}
It is easy to find the co-product $\Delta_{\star}(\delta)$ which is compatible with $\star$-multiplication algebra:
\begin{equation}
\Delta_{\star}={\cal T}^{-1}\Delta_{0}{\cal T}~,
\label{16}
\end{equation}
where $\Delta_{0}(\delta)=\delta \otimes {\bf 1}+{\bf 1}\otimes \delta$ is the standard co-product compatible with the usual pointwise algebra. Under the twisted co-product (\ref{16}) we obtain that the modified Leibniz rule, i.e.
\begin{eqnarray}
\delta\left(\Phi_1\otimes \Phi_2\right)= \delta(\Phi_1)\Phi_2 - \Phi_1\delta(\Phi_2)\nonumber \\
+\sum_{n=1}^{\infty}\frac{(i\ell^2)^n}{2^nn!}\theta^{a_1b_1}...\theta^{a_nb_n}
\delta \left(A_{a_1}...A_{a_n},\Phi_1\right)\left(A_{b_1}...A_{b_n},\Phi_2\right) 
+\left(A_{a_1}...A_{a_n}\Phi_1\right)\delta \left( [ A_{b_1}...A_{b_n}\Phi_2\right)
\label{17}
\end{eqnarray}

At this point we would like to stress two important points of our present formalism. First, since the $\star$-product (\ref{6}) (and hence the twist operator (\ref{12})) involves covariant derivatives, there is no contradiction in eq. (\ref{17}), i.e. the product of fields is indeed transforms covariantly, and thus the criticism expressed in \cite{Alvarez-Gaume:2006bn} and in \cite{Chaichian:2006we}, \cite{Chaichian:2006wt} does not apply. Another important point is that, although the twist operator (\ref{12}) is non-Abelian, it acts as an Abelian twist in the space of space-time tensors. Indeed, it is easy to see that, the action of a commutator $[A_a,A_b]$ on a generic space-time tensor field $T$ which is singlet of SO(3.1) LLI group, is trivial, $[A_a,A_b]T=0$. This means that  the theory defined above is associative for space-tme tensors, unlike the theories discussed in \cite{Chaichian:2008pq} and also in \cite{Harikumar:2006xf}.

\section{The action}

Next we compute the objects needed to construct an invariant action under the GCI and twisted LLI symmetries described in the previous section. The covariant field strengths are defined through the $\star$-commutator:
\begin{eqnarray}
-i[A_{a},A_{b}]_{\star}=-i[A_{a},A_{b}]\nonumber \\
-i\sum_{n=1}^{\infty}
\frac{(-i\ell^2)^n}{n!2^n}\theta^{a_1b_1}...\theta^{a_nb_n}[A_{a_1},...,[A_{a_n}, A_a]...][A_{b_1},...,[A_{b_n}, A_b]]+i\sum_{n=1}^{\infty}\left(a\longleftrightarrow b\right),
\label{18}
\end{eqnarray} 
where 
\begin{equation}
-i[A_a,A_b]=\frac{1}{2}R_{ab}^{~~cd}\Sigma_{dc}+T_{ab}^{~~\nu}P_{\nu}~,
\label{19}
\end{equation}
\begin{equation}
R_{ab}^{~~cd}=\partial_{a}\omega_{b}^{~~cd}-\partial_{b}\omega_{a}^{~~cd} - \omega_{a}^{~~ce}\omega_{b~e}^{~~d}+\omega_{b}^{~~ce}\omega_{a~e}^{~~d}~,
\label{20}
\end{equation}
is the commutative Riemann curvature, and, 
\begin{equation}
T_{ab}^{~~c}\equiv T_{ab}^{~~\nu}e_{\nu}^{~~c}=\mathcal{C}_{ab}^{~~c}+\omega_{a~b}^{~~~c}-\omega_{b~a}^{~~~c}~,~~
\mathcal{C}_{ab}^{~~c}=\left(\partial_ae_{b}^{~\nu}-\partial_be_{a}^{~\nu}\right)e_{\nu}^{~c}
\label{21}
\end{equation} 
is the commutative torsion field. The torsion-free condition $T_{ab}^{~~c}=0$ can be solved straightforwardly to express the spin-connection field through the vierbein and its derivatives: 
\begin{equation}
\omega_{a~bc}=\frac{1}{2}\left(\mathcal{C}_{cb~a}-\mathcal{C}_{ab~c}+\mathcal{C}_{ba~c}\right)~.
\label{22}
\end{equation}
The noncommutative Riemann tensor, $\mathcal{R}_{ab}^{~~cd}$, then can be calculated from (\ref{18}) as:
\begin{eqnarray}
\mathcal{R}_{ab}^{~~cd}=-i\mathrm{Tr}\left([A_{a},A_{b}]_{\star}\Sigma^{cd}\right)\nonumber \\
=R_{ab}^{~~cd}+i\mathrm{Tr}\left(\sum_{n=1}^{\infty}
\frac{\ell^{4n}}{(2n)!4^n}\theta^{a_1b_1}...\theta^{a_{2n}b_{2n}}[~ 
[A_{a_1},...,[A_{a_{2n}}, A_a]...],[A_{b_1},...,[A_{b_{2n}}, A_b]...]~ ] \Sigma^{cd}\right)~.
\label{23}
\end{eqnarray}
The first term in the second line of the above equation is the commutative Riemann curvature, while the remaining terms are noncommutative corrections to it. These corrections comprise of commutative Riemann curvatures and their covariant derivatives. Noncommutative corrections contain only even number of $\theta$'s and thus are real. The terms with odd number of $\theta$'s drop out because they are proportational to $\mathrm{Tr}\left(\left\lbrace \Sigma,\Sigma\right\rbrace \Sigma \right)$, which is identically 0. This seems to be a common feauture of all hermitian theories of noncommutative gravity \cite{Kobakhidze:2007jn}. The invariant action then can be written as:
\begin{equation}
S_{\mathrm{NC~GR}}=\frac{1}{2\kappa^2}\int d^4x \mathrm{det}(e_{\mu}^{~a})\mathcal{R}~,
\label{24}
\end{equation} 
where $\mathcal{R}$ is the noncommutative Ricci scalar, $\mathcal{R}=\mathcal{R}_{ab}^{~~ab}$, and $\kappa^2=8\pi G_N$. This action is remarkably simpler than the actions for different versions of noncommutative gravities proposed so far. Therefore, studies of gravitational effects of spacetime noncommutativity, hopefully, will be easier to perform. We plan such investigations in future publications. 

\section{Conclusion and outlook}

In this paper we have succeeded, for the first time, in constructing diffeomorphism-invariant and associative theory of noncommutative gravity. The local Lorentz invariance is broken, but it is realized as twisted gauge symmetry. Besides maintaining the full diffeomorphism invariance, the action (\ref{24}) for our theory is remarkably simpler than other actions within the previous formulations of noncommutative gravity. Therefore, we hope that the effects of space-time noncommutativity, e.g., in cosmology and black hole physics can be investigated in more consistent way. 

The theory of noncommutative gravity necessarily contains higher-derivative terms, which might lead to perturbative renormalizability of the theory. This is impossible within the previous formulations, since the linearized graviton propagator in those models is not affected by the space-time noncommutativity \cite{Kobakhidze:2007jn}. Recently, a renormalizable higher derivative theory with broken diffeomorphism invariance has been proposed in \cite{Horava:2009uw}. We stress here that, to remove ghost states from the spectrum of the theory it is not necessary to break the diffeomorphism invariance. Similar effect can be in principle achieved by breaking local Lorentz invariance. Moreover, keeping the full diffeomorphism invariance might be necessary in view of the recent criticism in \cite{Li:2009bg}-\cite{Kobakhidze:2009zr}, according to which the theory suggested in \cite{Horava:2009uw} contains an  extra propagating mode. It remains to be seen whether the theory of noncommutative gravity described here is remormalizable and ghost states are absent for the particular pattern of $\theta$ (e.g., spatial noncommutativity). We will study these issues elsewhere.

\subparagraph{Acknowledgments.}
This work was supported by the Australian Research Council



\end{document}